\documentclass[letterpaper, 10 pt, conference]{ieeeconf}
\IEEEoverridecommandlockouts
\usepackage{graphicx}
\usepackage{epstopdf}
\usepackage{todonotes}

\usepackage[hidelinks]{hyperref}
\usepackage{amsmath,amsfonts,amssymb}
\usepackage{mathtools}
\usepackage{siunitx}
\usepackage{bigints}
\usepackage{longtable,tabularx}
\usepackage{booktabs}

\newtheorem{remark}{Remark}
\newtheorem{prob}{Problem}

\title{\LARGE \bf On-line Motion Planning Using Bernstein Polynomials for Enhanced Target Localization in Autonomous Vehicles\thanks{This research was supported by the Office of Naval Research, grant N000142112091}}
\author{Camilla Tabasso, Venanzio Cichella\thanks{C. Tabasso and V. Cichella are with the University of Iowa, Department of Mechanical Engineering, 52240, Iowa City, IA, email: {\tt\footnotesize \{camilla-tabasso,\, venanzio-cichella\}@uiowa.edu}}}
\date{May 2022}

\begin{document}
\thispagestyle{empty}
\pagestyle{empty}
\maketitle

\begin{abstract}
The use of autonomous vehicles for target localization in modern applications has emphasized their superior efficiency, improved safety, and cost advantages over human-operated methods. For localization tasks, autonomous vehicles can be used to increase efficiency and ensure that the target is localized as quickly and precisely as possible. However, devising a motion planning scheme to achieve these objectives in a computationally efficient manner suitable for real-time implementation is not straightforward. In this paper, we introduce a motion planning solution for enhanced target localization, leveraging Bernstein polynomial basis functions to approximate the probability distribution of the target's trajectory. This allows us to derive estimation performance criteria which are used by the motion planner to enhance the estimator efficacy. To conclude, we present simulation results that validate the effectiveness of the suggested algorithm.
\end{abstract}

\section{Introduction}
The deployment of autonomous vehicles for target localization has gained significant traction in modern applications, highlighting their enhanced efficiency, safety, and cost-effectiveness compared to human-operated localization. Prominent applications encompass search and rescue operations, where unmanned aerial vehicles play a crucial role in victim identification \cite{albanese2022sardo,sampedro2019fully}; surveillance and monitoring, spanning both defense and civil sectors \cite{pack2009cooperative}; industrial environments where robotic agents equipped with sensors to detect RFID-tagged items have revolutionized resource management processes \cite{liu2021accurate,buffi2018measuring}.

In numerous scenarios, it is imperative to ensure that the localization tasks are carried out in a \textit{timely} and \textit{accurate} fashion. Such prerequisites drive the development of control solutions to enhance the efficiency of the localization process. A widely studied methodology hinges on determining the optimal positioning of sensors to monitor the target. Notably, the Cramer-Rao Lower Bound (CRLB), a measure that sets a minimum threshold on the variance of an estimator, can facilitate the derivation of an explicit solution for the optimal geometry between sensor and target \cite{yang2012performance,sadeghi2020optimal,sheng2019optimal}. 
When deploying autonomous vehicles equipped with onboard sensors for estimation tasks, the dynamic behavior of these mobile sensors requires a shift from static optimal geometries to real-time optimal motion planning strategies. This shift promises improvements in localization's efficiency and accuracy, but also emphasizes the need to formulate a motion planning algorithm that should:
(1) integrate the localization process to optimize its effectiveness; (2) incorporate standard constraints inherent to motion planning algorithms for autonomous systems, such as collision avoidance, compliance with vehicle dynamics, feasibility constraints, among others; (3) exhibit computational efficiency, enabling quick re-planning as new information about the target becomes available. 

Several solutions to on-line motion planning for estimation and target localization have been proposed, which rely on single or multiple vehicles to perform localization. The use of multiple vehicles offers advantages for fast coverage of larger areas. However, multi-vehicle systems can be very costly and require long setup and deployment times, which is undesirable in time-critical applications. This motivates the need to develop robust and reliable solutions for localization scenarios where the use of a single vehicle is preferable.

From the perspective of motion planning, the localization problem is more complex when undertaken by a single agent \cite{tabasso2021Optimal}. This is because generating "sufficiently rich" trajectories becomes even more challenging. Nevertheless, several works that consider single-vehicle motion planning for efficient localization can be found in the literature \cite{anjaly2018observability,crasta2018multiple,roh2018trajectory,hung2020range}. In \cite{anjaly2018observability}, a method based on maximizing the observability of a maneuvering target given relative bearing measurements is analyzed. In \cite{crasta2018multiple}, the authors formulate an optimal trajectory generation algorithm that maximizes the parametric Fisher Information Matrix (FIM) for a localization scenario. In \cite{roh2018trajectory}, a similar problem is solved by generating optimal trajectories that minimize the weighted sum of CRLB diagonal entries. One drawback of these methods is that the trajectories they generate only optimize FIM related costs for achieving time-efficient localization. However, a more complete approach to efficient target localization may also include other objectives and constraints, such as energy consumption, terminal costs and safety requirements. To deal with this limitation, in \cite{hung2020range}, the authors tackle the problem of multi-objective motion planning for localization and pursuit of moving targets. 
The problem is solved by designing an MPC-based controller that maximizes the predicted Bayesian Fisher Information Matrix. However, in MPC-based motion planning, it is often challenging to find a suitable discretization scheme that is able to guarantee both computational efficiency required for real-time implementation and accuracy in the solution.

With these ideas in mind, we propose a solution to motion planning for enhanced target localization based on the use of Bernstein polynomial basis functions for the approximation of both vehicle's trajectory and target's probability distribution. In particular, we formulate the problem at hand as an optimal control problem. We employ a direct method based on Bernstein polynomials that was initially proposed in \cite{cichella2017optimal}. Bernstein polynomials posses geometric properties that are particularly useful for the computation and enforcement of feasibility and safety constraints. 
Bernstein polynomials have also attracted a lot of attention as a mean to estimate distribution functions since they are known to yield smooth estimates and do not suffer from the Gibbs phenomenon when estimating discontinuous functions, see \cite{vitale1975bernstein,tenbusch1994two,babu2002application,leblanc2012estimating} and references therein. 
Therefore, the main novelty of this work is the development of a flexible and efficient motion planning framework for localization tasks. The use of Bernstein polynomials allows us to quickly and efficiently estimate the location of a target by integrating estimation performance criteria, in terms of efficiency and accuracy, into the motion planner. This, in turn, results in a decrease in the total time needed to locate the target.

This paper is organized as follows: the on-line motion planning problem for enhanced target localization is formally stated in Section \ref{sec:problemformulation}. Section \ref{sec:solution} presents the approximation method based on Bernstein polynomials used to solve the problem. Finally, numerical results are shown in Section \ref{sec:results}, while conclusions are presented in Section \ref{sec:conclusions}.

\section{Problem Formulation} \label{sec:problemformulation}
Consider an autonomous agent
positioned at $p(t) = [x(t) , y(t)]^\top$.  We assume that the agent is equipped with an autopilot and a trajectory tracking algorithm such that the generalized trajectory tracking error
\(e_p = || p(t) - p_d(t) ||\)
converges to and remains within a neighborhood of zero, where \(p_d= [x_d(t) , y_d(t)]\) is a desired trajectory satisfying 
\begin{equation} \label{eq:asmautopilot} 
||\dot{p}_d(t)|| \leq v_{\text{max}} , \qquad ||\ddot{p}_d(t)|| \leq a_{\text{max}},
\end{equation}
\noindent where $v_{\text{max}}$ and $a_{\text{max}}$ are the maximum speed and acceleration, respectively.
In the equations mentioned above, the norm \(||\cdot||\) is assumed to be the Euclidean norm. That is, for a vector \(x = [x_1, \ldots, x_n]\), the norm is computed as \(||x|| = \sqrt{x_1^2 + \ldots + x_n^2}\). Consider a stationary target positioned at \( p^{\star} \in \mathbb{R}^2 \). Assume both the target and the agent are equipped with sensors, allowing the agent to obtain estimates of the target's position, \( \hat{p}_i = [\hat{x}_i,\hat{y}_i]^\top \), at discrete time instants \( t_i, \quad i \in \{1,2,\ldots\} \). One practical implementation of such a localization system can be observed in avalanche situations: the target, representing a victim, and the agent, a drone, are both equipped with beacons. The beacon attached to the target provides Received Signal Strength (RSS) measurements, representing the relative signal strength of the beacon detected by a receiver. The measurement acquired by the receiver is used to determine the range between the beacon and the receiver, which is subsequently employed to estimate the target's position using, for example, filtering or trilateration techniques \cite{shi2020rssi,jondhale2022trilateration}. At discrete time intervals defined by \( t_k = k \Delta T \) where \( k \in \{1,2,\ldots\} \) and \(\Delta T\) is the time interval between re-planning events, the motion planner utilizes all prior estimates to update the vehicle's trajectory. This updated trajectory takes into account localization efficiency, accuracy, and the constraints of the vehicle.

Primarily, the motion planner must consider feasibility constraints to ensure that the vehicle can track the generated trajectory. These constraints encompass: \textit{dynamic feasibility} constraints, as described by Equation \eqref{eq:asmautopilot}; 
\textit{collision avoidance} constraints, ensuring the vehicle avoids obstacles. Given \(p_{\text{o},j}\) as the position of obstacle \(j\), the desired trajectory must meet:
    \begin{equation} \label{eq:collisionavoidance}
    ||p_d(t) - p_{\text{o},j}|| \geq d_{\text{safe}} , \qquad \forall t \geq 0, \quad \forall j \in \{1,\ldots,n_o\} ,
    \end{equation}
    where \(d_{\text{safe}}\) denotes the safety distance, and \(n_o\) represents the number of obstacles; and \textit{boundary conditions}, which require the initial position and velocity of the trajectory to match the vehicle's current state. Formally, this is represented as:
    \begin{equation} \label{eq:boundaryconditions}
    p_d (t_i) = p(t_i),  \quad \dot{p}_d (t_i) = \dot{p} (t_i).
    \end{equation}

In order for the motion planning algorithm to account for localization efficiency, we take inspiration from \cite{oshman1999optimization,ponda2009trajectory} where the determinant of the FIM is used as a performance measure for the estimation process. The determinant of the FIM at time $t_i$ computed over the desired trajectory assigned to the vehicle is defined as follows:
\begin{equation} \label{eq:fim}
\det \mathcal{I}_i(p_d(t)) = \det \frac{1}{\sigma^2} \sum_{k=i}^{n}
    \begin{bmatrix}
    \frac{\Delta x_k^2}{r_k^2} & -\frac{\Delta y_k \Delta x_k}{r_k^2} \vspace{3mm} \\
    -\frac{\Delta y_k \Delta x_k}{r_k^2} & \frac{\Delta y_k^2}{r_k^2}
    \end{bmatrix} ,
\end{equation}
where $\sigma^2$ is the variance of the sensor noise and 
\begin{equation} \label{eq:FIMterms}
\begin{gathered} 
\Delta x_k = x_d(t_k) - \hat{x}_i, \quad \Delta y_k = y_d(t_k) - \hat{y}_i, \\
r_k = \sqrt{\Delta x_k^2 + \Delta y_k^2}. 
\end{gathered}
\end{equation}
We notice that $\mathcal{I}_i(p_d(t))$ depends on the geometry of the vehicle's path. We can leverage this insight by designing a cost function that uses the determinant of the FIM to improve the performance of the estimator.

Finally, it is crucial for motion planning to guide the agent towards the target's location for several reasons. Firstly, range measurements are influenced by disturbances, which decrease as the vehicle approaches the target. Secondly, in emergency situations, it may be imperative for the agent to be in close proximity to the target or victim. However, the target location is not known, and only estimates at discrete times are available. Then, we need to devise a cost that steers the vehicle toward a particular location in the search area which is most likely to be the true location of the target. Let \( f(\zeta) = [f_x(\zeta_x), f_y(\zeta_y)]^\top\) denote the probability density function (PDF) describing the position of the target. To maximize the likelihood of the agent reaching the target at a final time \( t_f \), one should maximize \( ||f(p_d(t_f))|| \), i.e. the final point of the desired trajectory assigned to the agent should be where the target is most likely to be located. 

With these considerations in mind, we define the motion planning problem for efficient localization as follows.
\begin{prob} \label{prob:problemOCP}
	At time $t_i$ compute final time \(t_f\) and trajectory $p_d:[t_i,\,t_f] \to \mathbb{R}^2$ that minimizes
		\begin{equation}
			\begin{split} 
				J =&  w_1(t_f-t_i) +w_2 \displaystyle{ \int_{t_i}^{t_f}} || \ddot{p}_d(\tau) ||^2\,d\tau  \\
				& - w_4 \det \mathcal{I}_i(p_d(t)) - w_3 ||f(p_d(t_f))|| 
			\end{split} 
		\end{equation}
        with $w_1, w_2, w_3, w_4 > 0$, subject to Equations \eqref{eq:asmautopilot}-\eqref{eq:collisionavoidance}.
\end{prob}
\vspace{3mm}

\begin{remark}
The use of the pdf to describe the position of the target effectively acts as a weight. If the location of the target is quite uncertain, the maximization of \( ||f(p_d(t_f))|| \) is not very relevant because the function does not have peaks at particular values, and the maximization of other costs are prioritized. Instead, when the estimate has high probability, the vehicle is steered more aggressively toward that position to precisely pinpoint the target's exact location. 
\end{remark}

\begin{remark}
    We note that the Fisher Information Matrix presented in Equation \eqref{eq:fim} is derived based on range measurements. However, the framework presented in this paper can be applied to vehicles with various types of on-board sensors. In this case, the FIM can be reformulated based on a different measurement model, i.e., $\mathcal{I} = \frac{1}{\sigma^2} \sum_{k=i}^{n} H_k^\top H_k$, where $H_k$ is the Jacobian of the measurement model.
\end{remark}

Given the complexity of Problem \ref{prob:problemOCP}, numerical methods need to be deployed to find optimal trajectory $p_d^* : [t_i,t_f]\to\mathbb{R}^2$ that solves it. 
In the subsequent section, we introduce an approximation method that transcribes Problem \ref{prob:problemOCP} into a non-linear programming problem, which can then be tackled by off-the-shelf optimization software.

\section{Bernstein Approximation of Problem \ref{prob:problemOCP}}
\label{sec:solution}
Let the desired trajectory assigned to the agent, \( p_d(t)\), be parameterized by the 2D $m$th order Bernstein polynomial
\begin{equation} 
    \label{eq:bernpolypd}
	\begin{split} 
		p_d(t) \triangleq p_{m}(t) & = \bar{c}_{m}^\top b_{m}(t),\qquad t \in [t_i,t_f], \\ 
	\end{split} 
\end{equation}
where $\bar{c}_{m} = [\bar{c}_{0,m},\ldots,\bar{c}_{m,m}]^\top \in \mathbb{R}^{(m + 1) \times 2}$
is the vector of Bernstein polynomial coefficients, \(b_{m}(t) = [b_{0,m}(t),\ldots,b_{m,m}(t)]^\top \in \mathbb{R}^{m+1} \) with  
\begin{equation}
    b_{j,m}(t) = \binom{m}{j} \frac{(t-t_i)^j(t_f-t)^{m-j}}{(t_f-t_i)^m}  \, , \qquad t \in [t_i,t_f], \label{eq:Bernsteinbasis}
\end{equation}
for $j=0,\ldots,m$, and 
$
\binom{m}{j} = \frac{m!}{j!(m-j)!}
$. In what follows, we demonstrate how to approximate the constraints and cost functions of Problem \ref{prob:problemOCP} by leveraging specific properties of Bernstein polynomials. 

\subsection{Inequality constraints}
The derivative of $p_{m}(t)$ is given by
	\begin{equation} \label{eq:derivatives}
			\dot{p}_{m}(t)  = \bar{c}_{m}^\top {D}_{m} b_{m-1}(t) , 
	\end{equation}
where \(D_m\) is the differentiation matrix given by
 \begin{equation}
            D_m =   \frac{m}{t_f - t_i} \begin{bmatrix}
                -1    & 0      & \cdots & 0       \\
                1      & \ddots & \ddots & \vdots  \\
                0      & \ddots & \ddots & 0       \\
                \vdots & \ddots & \ddots & -1      \\
                0      & \cdots & 0      & 1
            \end{bmatrix} \in \mathbb{R}^{m+1 \times m}.
        \end{equation}

The sum (difference) of two $m$-th order Bernstein polynomials can be computed by adding (subtracting) their  Bernstein coefficients. Also, the product between two Bernstein polynomials of orders $m$ and $n$, with coefficients $\bar{f}_{m}=[\bar{f}_{0,m}, \dots, \bar{f}_{m,m}]$ and $\bar{g}_{n}=[\bar{g}_{0,n}, \dots, \bar{g}_{n,n}]$, respectively, is a Bernstein polynomial of order $m+n$ with coefficients given by 
	\begin{equation*}
	    \bar{c}_{k,m+n} = \sum_{j=\max(0,k-n)}^{\min(m,k)} \frac{\Big( \begin{array}{c} {m} \\ {j} \end{array} \Big) \Big( \begin{array}{c} {n} \\ {k-j} \end{array} \Big) }{\Big( \begin{array}{c} {m+n} \\ {k} \end{array} \Big) } \bar{f}_{j, m} \hspace{1mm} \bar{g}_{k-j, n}.
	\end{equation*}
By applying differentiation and arithmetic operations to Bernstein polynomials, the expressions on the left-hand sides of Equations \eqref{eq:asmautopilot} and \eqref{eq:collisionavoidance} can also be represented as Bernstein polynomials:
\begin{equation} \label{eq:norms}
	\begin{split}
	    ||\dot{p}_{m}(t)||^2  = \bar{v}_{2m}^\top b_{2m}(t), \\
	    ||\ddot{p}_{m}(t)||^2  = \bar{a}_{2m}^\top b_{2m}(t), \\
	    ||{p}_{m}(t)-p||^2 = \bar{d}_{2m}^\top b_{2m}(t).
	\end{split}
\end{equation}
The coefficients $\bar{v}_{j,m}$, $\bar{a}_{j,m}$ and $\bar{d}_{j,m}$, $\forall j \in \{0,\ldots, 2m \}$ can be obtained from algebraic manipulation of the Bernstein coefficients of ${p}_{m}(t)$, \(\bar{c}_m\).

To enforce the inequality constraints, Equations \eqref{eq:asmautopilot} and \eqref{eq:collisionavoidance}, we use the convex hull property of Bernstein polynomials: they are bounded by the convex hull of their control points. This property implies that \[\min_i \bar{c}_{i,m} \leq p_m(t) \leq \max_i \bar{c}_{i,m} . \] 
The property is useful to enforce min and max constraints by imposing the same onto the control points of the polynomials. However, the convex hull is often quite a bit larger than the curve itself, and is not a ``tight'' fit, leading to inefficiencies. To reduce this conservativeness, degree elevation can be used. Any Bernstein polynomial of degree $m$ can be expressed as a Bernstein polynomial of degree $m_{el}$, $m_{el} > m$. The vector of coefficients of the degree elevated Bernstein polynomial, $\bar{c}_{m_{el}}$, can be calculated as \(\bar{c}_{m_{el}} = {E}_{m,m_{el}}^\top \bar{c}_{m} ,\)
where ${E}_{m,m_{el}} = \{e_{j,k}\} \in \mathbb{R}^{(m+1) \times (m_{el}+1)}$ is the degree elevation matrix with elements given by
        \begin{equation} \label{eq:degelevmatrix}
            e_{i, i+j} = \frac{\binom{m_{el}-m}{j} \binom{m}{i}}{\binom{m_{el}}{i+j}} \, ,
        \end{equation}
        where $i = 0, \dots, m$ and $j = 0, \dots, m_{el} - m$, all other values in the matrix are zero (see \cite{lee1997distance}).
        Because ${E}_{m,m_{el}}$ is independent of the coefficients, it can be computed ahead of time and stored for efficient computation of degree elevated Bernstein polynomials.
        When a Bernstein polynomial undergoes degree elevation, its coefficients converge to the polynomial. Intuitively, this is because the control points of a polynomial elevated in degree are convex combinations of the control points from the original polynomial. For theoretical insights on the convergence of these control points to the polynomial during degree elevation, refer to \cite{prautzsch1994convergence}.
With this is mind, we employ the degree elevation property to ensure dynamic feasibility and adherence to collision avoidance constraints. We rewrite the expressions in Equation \eqref{eq:norms} as follows:
\begin{equation} \label{eq:normselev}
	\begin{split}
	    ||\dot{p}_{m}(t)||^2  = \bar{v}_{2m}^\top{E}_{2m,m_{el}} b_{m_{el}}(t) = \bar{v}_{m_{el}}^\top b_{m_{el}}(t) , \\
	    ||\ddot{p}_{m}(t)||^2  = \bar{a}_{2m}^\top{E}_{2m,m_{el}} b_{m_{el}}(t)
     = \bar{a}_{m_{el}}^\top b_{m_{el}}(t) , \\
	    ||{p}_{m}(t)-p||^2 = \bar{d}_{2m}^\top{E}_{2m,m_{el}} b_{m_{el}}(t)
     = \bar{d}_{m_{el}}^\top b_{m_{el}}(t) ,
	\end{split}
\end{equation}
with $m_{el} > 2m$ and sufficiently large. Then, constraints \eqref{eq:asmautopilot} and \eqref{eq:collisionavoidance} can be imposed as follows:
\begin{equation} \label{eq:feasandcoll_disc}
\begin{gathered}
\bar{v}_{i,m_{el}} \leq v_{\max} , \qquad \bar{a}_{i,m_{el}} \leq a_{\max} \\
\bar{d}_{i,m_{el}} \geq d_{\text{safe}}, \qquad \forall i \in \{0 , \ldots, m_{el}\}
\end{gathered}
\end{equation}

\subsection{Boundary conditions}
The first and last coefficients of Bernstein polynomials are their endpoints, i.e.,
\({p}_d(t_i) = \bar{c}_{0,m}\) and \({p}_d(t_f) = \bar{c}_{m,m}\).
By combining this with Equation \eqref{eq:derivatives}, we can express the boundary conditions \eqref{eq:boundaryconditions} in the form:
\begin{equation} \label{eq:boundaryconditions_disc}
\bar{c}_{0,m} = p(t_i), \qquad \frac{m}{t_f-t_i} (\bar{c}_{1,m}-\bar{c}_{0,m}) = \dot{p}(t_i) .
\end{equation}

\subsection{Cost function}
The definite integral of Bernstein polynomial can be computed as the weighted sum of its control points. I.e., the definite integral of $||\ddot{p}_{d}(t)||^2$ can be computed as
	\begin{equation} \label{eq:integralbezier}
		\int_{t_i}^{t_f}||\ddot{p}_{d}(\tau)||^2d\tau = \frac{t_f-t_i}{m+1}  \sum_{j = 0}^{m} \bar{a}_{j,m} \, ,\, .
	\end{equation}

To calculate the determinant of the FIM, the de Casteljau algorithm is utilized. The de Casteljau algorithm offers a numerically stable approach to evaluate a curve at a given $t$.
The subsequent steps detail the application 
of the de Casteljau algorithm to compute \( p_d(t_k) \):
\begin{enumerate}
    \item For \( i = 0, \ldots, m \), compute:
    \[ \bar{c}_{i,m}^{(0)} = \bar{c}_{i,m}. \]
    \item For \( j = 1, \ldots, m \) and \( i = 0, \ldots, m-j \), compute:
    \[ \bar{c}_{i,m}^{(j)} = \frac{t_f-t_k}{t_f-t_i} \bar{c}_{i,m}^{(j-1)} + \frac{t_k-t_i}{t_f-t_i} \bar{c}_{i+1,m}^{(j-1)}. \]
    \item The curve point is given by:
    \[ p_d(t_k) = \bar{c}_0^{(m)}. \]
\end{enumerate}
With the above steps, the determinant of the FIM in Equation \eqref{eq:fim} can be computed as
\begin{equation}
\det \mathcal{I}_i (p_d(t)) = \sum_{k=i}^n \sum_{j>k}^n \frac{\left( \Delta x_k\Delta y_j - \Delta x_j \Delta y_k \right)^2}{r_k^4 r_j^4},
\end{equation}
with \(\Delta x_k, \Delta y_k\) and \(r_k\) defined in Equation \eqref{eq:FIMterms}.

Regarding the approximation of the PDF of the target's position, the procedure is outlined as follows: 
given the incoming estimates, the empirical cumulative distribution function (ECDF) is computed. 
Subsequently, this ECDF is approximated using Bernstein polynomial approximation, 
resulting in a Bernstein polynomial that approximates the actual cumulative distribution function (CDF) of the target's position. 
By differentiating this Bernstein polynomial, an approximation of the PDF of the target's position is obtained. 
Bernstein polynomials are advantageous for estimating CDF due to their ability to act as a stable filter when approximating the CDF from the ECDF, ensuring a more accurate representation. They exhibit remarkable stability when approximating non-smooth functions, avoiding the Gibbs phenomenon, and their derivatives converge, making them a reliable choice for such approximations.

Let $\{\hat{p}_{1}, \cdots \hat{p}_{n} \}$ be the set of estimates that we obtain at discrete time steps $\{t_{1},\ldots,t_n\}$. The ECDF of the target's location, $F_n(\zeta)$, $\zeta \triangleq [\zeta_x, \zeta_y] \in Z \subseteq \mathbb{R}^2$, $\zeta_x, \zeta_y \in [\zeta_{\min}, \zeta_{\max}]$, is given by
\begin{equation} \label{eq:empiricalCDF}
    F_n(\zeta) = \begin{bmatrix} F_n^{(x)}(\zeta_x) \\ F_n^{(y)}(\zeta_y) \end{bmatrix}  =
    \begin{bmatrix}
    \frac{1}{n} \sum_{i=1}^n I ( \hat{x}_i \leq \zeta_x) \\[3pt]
    \frac{1}{n} \sum_{i=1}^n I ( \hat{y}_i \leq \zeta_y)
    \end{bmatrix} ,
\end{equation}
where $I(a \leq b)$ yields $1$ when $a \leq b$, and $0$ otherwise. Then, the $m$-th order Bernstein estimator of the CDF, \(F(\zeta)\), is the Bernstein polynomial approximation of the ECDF, i.e.,
\begin{equation} \label{eq:BernsteinEstimator}
    {F}_{m,n}(\zeta) =
    \left[ 
    \begin{array}{c}
     \left(\bar{F}_{m,n}^{(x)}\right)^\top  b_{m} (\zeta_x)  \\
     \left(\bar{F}_{m,n}^{(y)}\right)^\top  b_{m} (\zeta_y) 
    \end{array}
    \right] , \quad 
       \zeta_x, \zeta_y   \in [\zeta_{\min}, \zeta_{\max}],
\end{equation}
with \( \bar{F}_{m,n}^{(x)} = [F_n^{(x)}(\zeta_{0,x}),\ldots,F_n^{(x)}(\zeta_{m,x})] \), \(\zeta_{i,x} = \frac{i}{m}(\zeta_{\max}-\zeta_{\min}) +\zeta_{\min}\) and similarly for \(\bar{F}_{m,n}^{(y)}\). 
Finally, the density function $f(\zeta)$ can be approximated by differentiating \eqref{eq:BernsteinEstimator}, that is,
 \begin{equation} \label{eq:f_hat}
    {f}_{m,n}(\zeta) = 
    \left[ \begin{array}{c}
     \left(\bar{F}_{m,n}^{(x)}\right)^\top D_m  b_{m-1} (\zeta_x)  \\
     \left(\bar{F}_{m,n}^{(y)}\right)^\top D_m  b_{m-1} (\zeta_y) 
    \end{array} \right].
\end{equation}
Using an argument similar to the one in \cite[Theorem 3.1]{babu2002application}, it can be shown that the Bernstein polynomial \(f_{m,n}(\zeta)\) converges to \(f(\zeta)\), i.e., 
\[||f_{m,n}(\zeta)-f(\zeta)||\to 0 \quad \text{a.s. as} \quad m,n \to \infty . \] 
Thus, the term \(||f(p_d(t_f))||\) appearing in the cost of the optimal motion planning problem can be approximated using \(||f_{m,n}(p_d(t_f))||\), which can be evaluated using the de Casteljau algorithm.

With this setup, the optimal motion planning problem can then be rewritten as the following nonlinear programming problem.

\begin{prob} \label{prob:problemNLP}
	At time $t_i$ compute final time \(t_f\) and vector of coefficients \(\bar{c}_m\), that minimize
		\begin{equation}
			\begin{split} 
				J =&  w_1(t_f-t_i) +w_2 \frac{t_f-t_i}{m+1} \sum_{j = 0}^{m} \bar{a}_{j,m}  \\
				-&w_3 ||f_{m,n}(p_d(t_f))|| - w_4 \det \mathcal{I}_i(p_d(t)),
			\end{split} 
		\end{equation}
        with $w_1, w_2, w_3, w_4 > 0$, subject to Equations \eqref{eq:feasandcoll_disc} and \eqref{eq:boundaryconditions_disc}.
\end{prob}

The solution to the aforementioned problem yields an optimal final mission time \( t_f^* \) and a vector of optimal Bernstein coefficients \(\bar{c}_m^*\). These are utilized to construct the Bernstein polynomial as defined in Equation \eqref{eq:bernpolypd}. As the problem is finite-dimensional, standard optimization software can be employed for its resolution.

\section{Implementation and numerical results}
\label{sec:results}
To validate the results in a realist setting, we assume that the agent is equipped with a receiver and the target is equipped with the transmitter. At each time step, the receiver obtains range measurements with respect to the transmitter which are degraded by Gaussian white noise, i.e., $n \sim \mathcal{N}(\mu,\sigma)$.
The range is then used to compute an estimate of the target's position using a least squares estimator implemented on MATLAB  $\mathtt{fminunc}$. 
 This estimator updates the estimate of the target's location, $\hat{p}_i$ at each time step as new readings are obtained.

The motion planning problem given in Problem \ref{prob:problemNLP} is solved using MATLAB $\mathtt{fmincon}$. We set $v_{max} = \SI{1}{m/s}$ and $a_{max} = \SI{1}{m/s}$. New target location estimates become available at a frequency of $\SI{2}{\hertz}$. The Bernstein estimator of the CDF is generated with order $m = \lceil n^{3/4}+2 \rceil$. The trajectories are generated with a replanning interval $\Delta T = \SI{5}{s}$, and termination criteria are set to $r_t = \SI{1}{m}$ and $t_{f,max} = \SI{150}{s}$. The average computational time for each trajectory replanning, i.e., the computational time needed to solve the optimal motion planning problem, was $0.05$ to $\SI{0.1}{s}$.

To demonstrate the efficacy of our method, we performed two simulations: the first one uses the motion planning problem given in Problem \ref{prob:problemNLP}, while the second simulation does not include the term in the cost function which maximizes the determinant of the FIM. The results of the two simulations are shown in blue and cyan, respectively. 

\begin{figure}
    \centering
    \includegraphics[width=8.0cm]{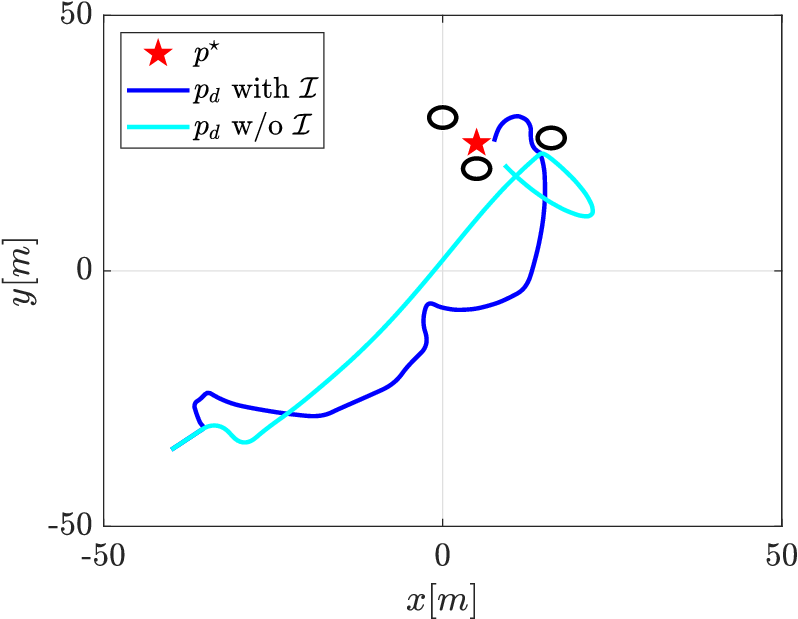}
    \caption{Path traveled by the vehicle during the mission.}
    \label{fig:traj}
\end{figure}

\begin{figure}
    \centering
    \includegraphics[width=8.0cm]{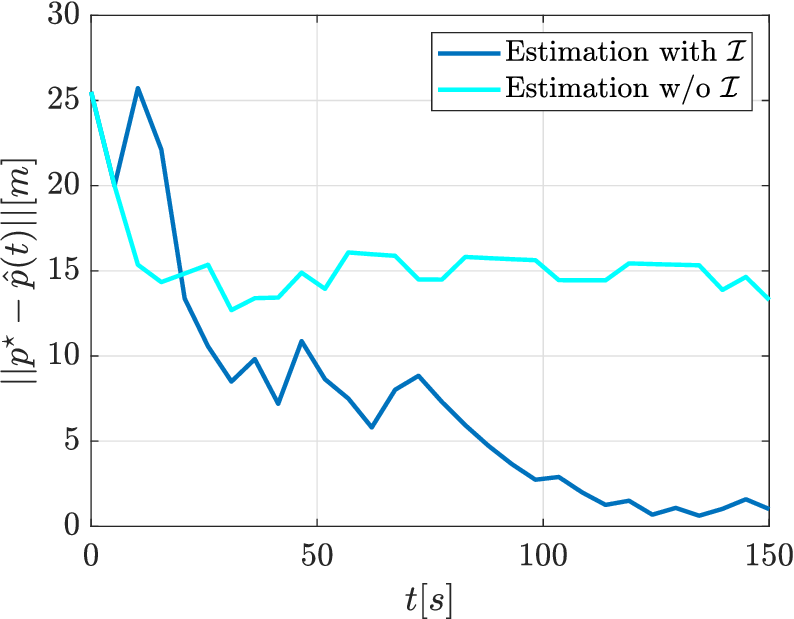}
    \caption{Estimation error over time.}
    \label{fig:estimation}
\end{figure}

The execution of both simulations is shown in Figure \ref{fig:traj}. It can be noticed that in the first simulation, the vehicle follows a longer path around the target, and by doing so is able to collect more meaningful measurements, which lead to better estimates. This can be seen more clearly in Figure \ref{fig:estimation}, which shows the estimation error over time. It can be noticed that throughout the entire simulation, this path results in a much smaller estimation error. On the other hand, in the second simulation, the vehicle quickly moves toward the estimated location of the target in a more linear fashion and subsequently circles around a small area. By doing so, the agent collects range measurements close to each other, which do not provide enough information to converge toward the true location of the target.

\begin{figure}
    \centering
    \includegraphics[width=8.0cm]{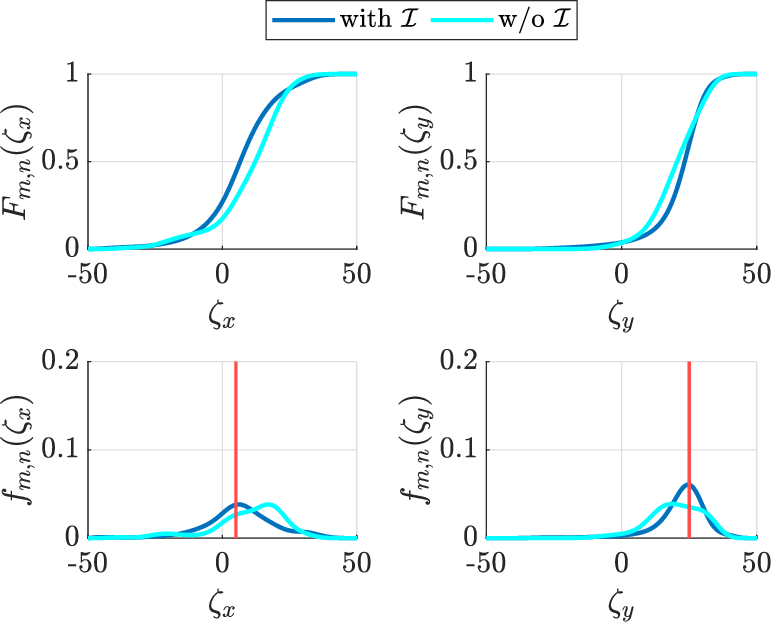}
    \caption{Cumulative density function (top row) for $x$ and $y$ coordinates, respectively, and probability density function (bottom row) for $x$ and $y$ coordinates, respectively. The true position of the target is represented by the red vertical line.}
    \label{fig:probabilities}
\end{figure}

Finally, Figure \ref{fig:probabilities} shows the CDF and PDF for both $x$ and $y$ coordinates at the time of the last trajectory replanning. The true location of the target is represented by the vertical red line. It can be seen that for the second simulation, the probability curves are skewed, and the density functions have a lower peak, which indicates less confidence in the estimates.

To further analyze the capabilities of our method, two more scenarios were studied, and for each one 25 simulations were conducted. In the first case, the position of the target was selected randomly in the search area i.e. $p(0) \sim [\mathcal{U}(-35,35), \mathcal{U}(-35,35)]^\top$, while the measurements were degraded by adding Gaussian noise with $\sigma = \SI{0.1}{m}$. In the second case, the position of the target was fixed, specifically at $p^{\star} = [-25, 15]^T$, while the standard deviation of the noise was variable, i.e., $\sigma ~ \sim \mathcal{U}(0,1).$ Both cases were repeated twice: once using the FIM in the motion planner and once without. For all the simulations, the trajectories were replanned with $\Delta T = \SI{5}{s}$ and the termination criteria were set to $r_t = \SI{1}{m}$ and $t_{f,max} = \SI{400}{s}$ to allow convergence of both methods. The results of the simulations were averaged and are showed in Table \ref{Tab:sim_results}, where the third column shows the final time required to successfully locate the target, while the fourth and fifth columns indicate the localization error, $e_{loc}(t) = ||p^{\star} - \hat{p}(t)||$, at $t = t_f$ and $t = \SI{20}{s}$, respectively.
 It can be seen that the simulations where the FIM was included in the cost function, greatly outperform their counterpart in terms of lower localization time. Moreover, it can be noticed that there is a faster reduction of the localization error when the FIM matrix is maximized, as shown in the last column of Table \ref{Tab:sim_results}, where the localization error after $\SI{20}{s}$ is roughly ten times lower.

\begin{table}
\begin{center}
\caption{Simulation results.}
\label{Tab:sim_results}
\begin{tabular}{ccccc}
  \toprule
  Sim. & $\sigma [m]$ & $t_f [s]$ & $e_{loc}(t_f) \, [m]$ & $ e_{loc}(20) \, [m]$ \\[5pt]
  \midrule
  w/o FIM & 0.1 & 237 & $<1$ & 21.7 \\ [5pt]
  \midrule
  with FIM & 0.1  & 197 & $<1$ & 2.7 \\[5pt]
  \midrule
  w/o FIM & $\sim \mathcal{U}(0,1)$ & 286.8 & $<1$ & 57.8 \\[5pt] 
  \midrule
  with FIM & $\sim \mathcal{U}(0,1)$ & 212.7 & $<1$ & 4.9 \\[5pt]
  \bottomrule
\end{tabular} 
\end{center}
\end{table}

\section{Conclusions} \label{sec:conclusions}


This paper presented a novel motion planning solution for target localization using autonomous vehicles. We leveraged Bernstein polynomial basis functions and highlighted the potential of Bernstein polynomials in optimizing autonomous vehicle trajectories for efficient target localization. Our approach, framed as a multi-objective non-linear optimal control problem, demonstrated enhanced estimator efficacy and addressed non-linear constraints inherent to autonomous vehicle missions. Simulations underscored the advantages of our method, especially when incorporating the Fisher Information Matrix in the cost function, resulting in faster and more accurate localization.  Future research will explore real-world applications and integrate advanced computational techniques for further refinement.

\bibliographystyle{ieeetr}
\bibliography{root}

\end{document}